\documentclass[a4paper,11pt]{article}
\pdfoutput=1 

\usepackage{jheppub} 

\usepackage{graphicx}
\usepackage{dcolumn}
\usepackage{bm}
\usepackage{amssymb}
\usepackage{amsmath}
\usepackage{epsfig}    
\usepackage{color}
\usepackage{slashed}
\usepackage{hhline}

\def\be{\begin{equation}}
\def\ee{\end{equation}}
\newcommand{\bea}{\begin{eqnarray}}
\newcommand{\eea}{\end{eqnarray}}

\numberwithin{equation}{section}

\title{$b\rightarrow s \mu^+ \mu^-$ anomalies and related phenomenology 
\\ in $U(1)_{B_3 - x_\mu L_\mu - x_\tau L_\tau}$ flavor gauge models}

\author[a,b]{ P. Ko}
\emailAdd{pko@kias.re.kr}
\affiliation[a]{School of Physics, KIAS, Seoul 02455, Korea}
\affiliation[b]{Quantum Universe Center, KIAS, Seoul 02455, Korea}

\author[a]{Takaaki Nomura}
\emailAdd{nomura@kias.re.kr}

\author[c]{Chaehyun Yu}
\emailAdd{chyu@korea.ac.kr}
\affiliation[c]{Department of Physics, Korea University, Anam-ro 145, 
Sungbuk-gu, Seoul 02841, Korea}

\abstract{We propose  a generation dependent lepton/baryon gauge symmetry, 
$U(1)_{B_3 - x_\mu L_\mu - x_\tau L_\tau} \equiv U(1)_X$   (with $x_\mu + x_\tau =1$ 
for anomaly cancellation),  as a possible solution for the $b\rightarrow s \mu^+ \mu^-$ anomalies.   
By  introducing two Higgs doublet fields, we can reproduce the observed  CKM matrix,  and 
generate flavor changing $Z'$ interactions in  the quark sector.  Thus one can explain observed  
anomalies in $b \to s \ell^+ \ell^-$ decay with the lepton non-universal $U(1)_X$ charge 
assignments. We show the minimal setup explaining $b \to s \ell^+ \ell^-$ anomalies, neutrino masses 
and mixings and dark matter candidate, taking into account experimental 
constraints of flavor physics such as charged lepton flavor violations and the 
$B_s$--$\bar B_s$ mixing.
Finally we  discuss collider physics focusing on $Z'$ production at the Large Hadron Collider and 
relic density of our dark matter candidate.
}

\begin{document} 
\maketitle
\flushbottom

\section{Introduction}

Although the standard model (SM) of particle physics is very successful we still do not have clear 
understanding of the physics regarding the flavors;  namely the origin of fermion masses and 
mixing patterns.
Then it is interesting to construct a model describing flavor physics with some symmetry 
as a guiding principle.  One of the attractive possibility is an introduction of flavor dependent 
$U(1)$ gauge symmetry which can constrain structure of Yukawa couplings generating masses 
for quarks, charged leptons and neutrinos.  In this kind of approaches to the flavor problem, 
these models may generate flavor changing neutral current (FCNC) processes through $Z'$ 
boson exchange, which will induce rich phenomenology.  

Recently there have been some indication of anomalies in $B$ physics measurements for 
$b \to s \ell^+ \ell^-$ process. 
The angular observable $P'_5$ in decay of $B$ meson, 
$B\to K^* \mu^+ \mu^-$~\cite{DescotesGenon:2012zf}, indicates $3.4\sigma$ deviations  
from the data with integrated luminosity of 3.0 fb$^{-1}$ at the LHCb~\cite{Aaij:2015oid}, confirming an earlier 
observation with $3.7\sigma$ deviations~\cite{Aaij:2013qta}. In addition, $2.1\sigma$ deviations were reported  
for the same observable by Belle~\cite{Abdesselam:2016llu, Wehle:2016yoi}. Furthermore, an anomaly in the measurement of lepton flavor universality by the ratio $R_K = BR(B^+ \to K^+ \mu^+\mu^-)/BR(B^+ \to K^+  e^+e^-)$~\cite{Hiller:2003js, Bobeth:2007dw} at the LHCb shows $2.6\sigma$ deviations from the SM prediction~\cite{Aaij:2014ora}.  
{ Moreover the LHCb collaboration also reported an anomaly in the ratio $R_{K^*} = BR(B \to K^* \mu^+\mu^-)/BR(B \to K^*  e^+e^-)$ where the observed values are deviated from the SM prediction by $\sim 2.4 \sigma$ as $R_{K^*} = 0.660^{+0.110}_{-0.070} \pm 0.024 (0.685^{+0.113}_{-0.069} \pm 0.047)$ for $(2 m_\mu^2) < q^2 < 1.1$ GeV$^2$ (1.1 GeV$^2 < q^2 < 6$ GeV$^2$)~\cite{Aaij:2017vbb}. }

These anomalies in the $b\rightarrow s \ell^+ \ell^-$ channels (with $\ell=e, \mu$) can be explained 
by flavor dependent $Z'$ interactions inducing effective operator 
of $(\bar b \gamma^\alpha s)(\bar \mu \gamma_\alpha \mu)$, if new physics contribution to the corresponding Wilson coefficient  $C_9^\mu$ is roughly $\Delta C_9^\mu \sim -1$ by 
  global fits~\cite{Capdevila:2017bsm, Altmannshofer:2017fio, Ciuchini:2017mik, Alok:2017sui}.
Then many models have been proposed to explain the anomalies by $Z'$ interactions~\cite{Crivellin:2015lwa, Sierra:2015fma, Ko:2017yrd,Bian:2017rpg, Bian:2017xzg, Duan:2018akc,Alonso:2017bff, Alonso:2017uky, Megias:2017ove, Boucenna:2016qad, Boucenna:2016wpr, Greljo:2015mma,Benavides:2018rgh, Hutauruk:2019crc, Chen:2017usq, Ko:2017quv, Geng:2018xzd, Baek:2019qte, Baek:2018aru, Darme:2018hqg, Faisel:2017glo, Chiang:2017hlj, Ko:2017lzd, Falkowski:2018dsl, Descotes-Genon:2017ptp, Baek:2017sew, Tang:2017gkz, DiChiara:2017cjq, Altmannshofer:2016jzy,Guadagnoli:2018ojc,Chala:2018igk}.

In this paper, motivated by $b \to  s \ell^+ \ell^-$ anomalies, we propose a model based on flavor 
dependent Abelian gauge symmetry $U(1)_{B_3 - x_\mu L_\mu - x_\tau L_\tau}$,  
which is anomaly-free for $x_\mu + x_\tau = 1$.
In this model we introduce two Higgs doublet fields to generate the realistic CKM matrix, 
where small mixings associated with third generation quarks can be obtained naturally as shown in Ref.~\cite{Crivellin:2015lwa}.
In the reference it is also shown that $Z'bs$ interaction is induced after electroweak symmetry breaking 
in a model with flavor dependent $U(1)_{L_\mu - L_\tau - a (B_1 + B_2 - 2 B_3)}$ gauge symmetry where $a$ can be arbitrary real number. 
Then, $b \to  s \ell^+ \ell^-$ anomalies can be explained by the effective operator induced by 
 exchange of a TeV scale $Z'$ boson. 
Following the same mechanism to induce $Z' b s$ interaction we can explain the anomalies by our flavor dependent $U(1)$ gauge symmetry
  if $x_\mu$ has negative value to get  $\Delta C_9^\mu \sim -1$.
We then consider the minimal model explaining $b \to  s \ell^+ \ell^-$ anomalies and generating 
non-zero neutrino masses in which two SM singlet scalar fields are introduced.  
Also we introduce Dirac fermionic 
dark matter (DM) candidate in order to account for the dark matter of the Universe.
In addition to $\Delta C_9^\mu$, we formulate neutrino mass matrix, lepton flavor violations (LFVs) 
and $B_s$--$\bar B_s$ mixing,  and experimental constraints from them are taken into account.
Then we discuss collider physics regarding $Z'$ production at the Large Hadron Collider (LHC) 
and relic density of our DM candidate.   

This paper is organized as follows. 
In Sec. II, we introduce our model and discuss quark mass, $\Delta C_9^\mu$ by $Z'$ and scalar 
masses  in the minimal case. 
In Sec.III we discuss neutrino mass matrix, charged lepton flavor violations and 
$B_s$--$\bar {B}_s$ mixing taking into account experimental constraints. 
The numerical analysis is  carried out in Sec. IV to discuss collider physics for $Z'$ production 
at the LHC and relic density of DM candidate showing allowed parameter region. 
Finally summary and discussion are given in Sec. V.

\section{ Models and formulas}

\begin{table}[t]
\begin{center} 
\begin{small}
\begin{tabular}{|c||c|c|c|c|c|c|c|c|c|c|c|c|c|c|c|c|}\hline\hline  
Fermions & $Q_{iL}$ & $u_{iR}$ & $d_{iR}$ & $Q_{3L}$ & $t_R$ & $b_R$ & 
$L_{1L}$ &$L_{2L}$ & $L_{3L}$ & $e_R$ & $\mu_{R}$ & $\tau_{R}$ & 
$\nu_{1R}$ & $\nu_{2R}$ & $\nu_{3R}$ 
\\\hline 
$SU(3)_C$ & $\bf{3}$ &  $\bf{3}$  & $\bf{3}$ & $\bf{3}$  & $\bf{3}$  & $\bf{3}$ & $\bf{1}$ & $\bf{1}$  & $\bf{1}$  & $\bf{1}$ & $\bf{1}$  & $\bf{1}$   & $\bf{1}$  & $\bf{1}$ & $\bf{1}$  \\\hline 
 $SU(2)_L$ & $\bf{2}$ & $\bf{1}$  & $\bf{1}$ & $\bf{2}$  & $\bf{1}$  & $\bf{1}$ & $\bf{2}$ & $\bf{2}$ & $\bf{2}$  & $\bf{1}$ & $\bf{1}$ & $\bf{1}$ & $\bf{1}$ & $\bf{1}$ & $\bf{1}$   \\\hline 
$U(1)_Y$ & $\frac16$ & $\frac23$  & $-\frac{1}{3}$ & $\frac16$ & $\frac23$  & $-\frac{1}{3}$ & $-\frac12$ & $-\frac12$  & $- \frac12$ & $-1$ & $-1$ & $-1$ & $0$ & $0$ & $0$    \\\hline
 $U(1)_{X}$ & $0$ & $0$ & $0$ & $\frac13$ & $\frac13$  & $\frac13$ & $0$ & $-x_\mu$  & $-x_\tau$ & $0$  & $-x_\mu$   & $-x_\tau$ & $0$  & $-x_\mu$ & $-x_\tau$   \\\hline
\end{tabular}
\caption{Charge assignment for the SM fermions and right-handed neutrinos where the indices 
$i = 1,2$ indicate the first and second generations.}
\label{tab:1}
 \end{small}
\end{center}
\end{table}

In this section we introduce our model based on flavor dependent 
$U(1)_{B_3 - x_\mu L_\mu - x_\tau L_\tau}$  gauge symmetry that we denote simply $U(1)_X$ 
in the following~\footnote{In our analysis we ignore 
kinetic mixing between $U(1)_Y$ and $U(1)_X$ assuming it is sufficiently small.}.
The SM fermions with 3 right-handed (RH) neutrinos are charged under the $U(1)_X$ as shown in 
Table.~\ref{tab:1}.  The gauge anomalies are cancelled when the $U(1)_X$ charges of fermions 
satisfy the condition
\begin{equation}
\label{Eq:anomaly}
x_\mu + x_\tau = 1 ,
\end{equation} 
which we will always assume in the following.  
In Sec. 2.1, we first discuss the case with general $x_{\mu, \tau}$ and investigate an explanation of 
$b \to  s \ell^+ \ell^-$ anomalies via flavor-changing $Z'$ interactions.
Then the minimal model with $x_\mu =  -1/3$ is constructed in Sec. 2.2,  taking into account the 
generation of active neutrino masses  and mixings  via Type-I seesaw mechanism.

\subsection{Discussion for general $(x_{\mu} , x_{\tau})$ case}

Firstly we consider quark sector which does not depend on our choice of $x_\mu$ and 
$x_\tau  = 1 - x_\mu$.   In this model we  have to introduce  at least  
two Higgs doublets  in order to induce the realistic  CKM mixing matrix:
\begin{equation}
\Phi_1 \ : \ ({\bf 1},{\bf 2})(1/2,-1/3), \quad \Phi_2 \ : \ ({\bf 1},{\bf 2})(1/2,0), \quad (SU(3)_C,SU(2)_L)(U(1)_Y, U(1)_X)
\end{equation}
Then the Yukawa couplings for quarks are given by 
\begin{align}
-{\cal L}_{Q} = & y^u_{i j}\bar Q_{i L}  \tilde\Phi_2 u_{j R} + y^d_{ij} \bar Q_{i L}\Phi_2 d_{j R} + y^u_{33}\bar Q_{3 L}  \tilde\Phi_2 t_R + y^d_{33} \bar Q_{3 L}\Phi_2 b_{R} \nonumber \\
& + \tilde y^u_{3i}\bar Q_{3 L}  \tilde\Phi_1 u_{i R} + \tilde y^d_{i3} \bar Q_{i L} \Phi_1 b_{R}  +{\rm h.c.},
\end{align}
where $i = 1,2$ and $\tilde \Phi_i = i \sigma_2 \Phi_i^*$.   
$\Phi_2$  is the Higgs doublet with vanishing  $U(1)_X$ charge, and is the SM-like Higgs doublet.
After two Higgs doublet fields get the non-zero vacuum expectation values (VEVs) 
$\langle \Phi_{1,2} \rangle =(0~ v_{1,2}/\sqrt{2})^T$, 
we obtain the following forms of quark mass matrices:
\begin{align}
& M^u = \frac{1}{\sqrt{2}} \left( \begin{array}{ccc} v_2 y^u_{11} & v_2 y^u_{12} & 0 \\ v_2 y^u_{21} & v_2 y^u_{22} & 0 \\ 0 & 0 & v_2 y^u_{33} \end{array} \right)
+ \left( \begin{array}{ccc} 0 & 0 & 0 \\ 0 & 0 & 0 \\  (\xi_u)_{31} & (\xi_u)_{32} & 0 \end{array} \right),  \nonumber \\
& M^d = \frac{1}{\sqrt{2}} \left( \begin{array}{ccc} v_2 y^d_{11} & v_2 y^d_{12} & 0 \\ v_2 y^d_{21} & v_2 y^d_{22} & 0 \\ 0 & 0 & v_2 y^d_{33} \end{array} \right)
+ \left( \begin{array}{ccc} 0 & 0 & (\xi_d)_{13} \\ 0 & 0 & (\xi_d)_{23} \\ 0 & 0 & 0 \end{array} \right).
\label{Eq:Massmatrix}
\end{align}
Note that  the matrices $\left(\xi_{u,d}\right)_{ij} \equiv \tilde y^{u,d}_{ij} v_1/\sqrt{2}$  have 
the same structure as those discussed in Ref.~\cite{Crivellin:2015lwa}.
We shall assume the second terms with $\xi_{u,d}$ are small perturbation effects generating 
realistic $3 \times 3$ CKM mixing matrix where the  $(33)$  elements are 
$v_2 y^{u(d)}_{33} \sim \sqrt{2} m_{t(b)}$ following the discussion in Ref.~\cite{Crivellin:2015lwa}.

As in the SM, the quark mass matrices are diagonalized by unitary matrices 
$U_{L, R}$ and $D_{L,R}$ which change quark fields from interaction basis to mass basis: 
$u_{L,R} \to U_{L,R}^\dagger u_{L,R} ~ (d_{L,R} \to D_{L,R}^\dagger d_{L,R})$. 
Then the CKM matrix is given by $V_{CKM} = U^\dagger_L D_L$.
Thus we obtain relation between mass matrices $M^{u,d}$ and diagonalized ones as follows:
\begin{equation}
M^d = D_L m^d_{\rm diag} D_R^\dagger, \quad M^u = U_L m^u_{\rm diag} U_R^\dagger,
\end{equation}
where diagonal mass matrices are given by $m^d_{\rm diag} = {\rm diag}(m_d,m_s,m_b)$ and $m^u_{\rm diag} = {\rm diag}(m_u,m_c,m_t)$.
Then $U_{L[R]}$ and $D_{L[R]}$ are associated with diagonalization of $M^u (M^u)^\dagger [ (M^u)^\dagger M^u]$ and $M^d (M^d)^\dagger [ (M^d)^\dagger M^d]$ by
\begin{align}
& M^{u} (M^{u})^\dagger \left[ (M^{u})^\dagger M^{u} \right] = U^\dagger_L (m^u_{\rm diag})^2 U_L \left[ U^\dagger_R (m^u_{\rm diag})^2 U_R \right], \nonumber \\
& M^{d} (M^{d})^\dagger \left[ (M^{d})^\dagger M^{d} \right] = D^\dagger_L (m^u_{\rm diag})^2 D_L \left[ D^\dagger_R (m^u_{\rm diag})^2 D_R \right].
\label{Eq:diagonalize}
\end{align} 
The structures of mass matrices in Eq.~(\ref{Eq:Massmatrix}) indicate that the off-diagonal elements associated with 3rd generations are more suppressed 
for $M^u (M^u)^\dagger$ and $(M^d)^\dagger M^d$ than those in $(M^u)^\dagger M^u$ and $M^d (M^d)^\dagger$. 
More specifically, we find that
\begin{align}
\left( M^u (M^u)^\dagger \right)_{31, 32, 13, 23} \left[ \left((M^d)^\dagger M^d \right)_{31, 32, 13, 23} \right] \propto  \frac{v_2}{\sqrt{2}} y_{ij} \xi_{3k[k3]}, \nonumber \\
\left( (M^u)^\dagger M^u \right)_{31, 32, 13, 23} \left[ \left(M^d (M^d)^\dagger \right)_{31, 32, 13, 23} \right] \propto  \frac{v_2}{\sqrt{2}} y_{33} \xi_{3k[k3]},
\end{align}
where $\{i, j, k \} = 1,2$.
Then we can approximate $U_L$ and $D_R$ to be close to unity matrix since they are 
 associated with diagonalizaition of $M^u (M^u)^\dagger$ and $(M^d)^\dagger M^d$, respectively, where mixing angles in $D_R(U_L)$ generated by $\xi$ parameters are suppressed by $m_{d,s(u,c)}/m_{b(t)}$ to those in $D_L(U_R)$. 
Therefore CKM matrix can be approximated as $V_{CKM} \simeq D_L$, and $D_R \simeq {\bf 1}$,  
as  obtained in Ref.~\cite{Crivellin:2015lwa}.  
Taking $D_L = V_{CKM}$, we can obtain sizes of $(\xi_d)_{13}$ and $(\xi_d)_{23}$  from Eq.~(\ref{Eq:diagonalize}) applying mass eigenvalues of down-type quarks.
We thus obtain
\begin{equation}
\left| (\xi_d)_{13} \right| \sim 0.034 \ {\rm GeV}, \quad \left| (\xi_d)_{23} \right| \sim 0.18 \ {\rm GeV}
\end{equation}
with $y_{33} v_2/\sqrt{2} \simeq m_b \simeq 4.2$ GeV.
Therefore we can reconstruct mass eigenvalues of down-type quarks with $D_L \simeq V_{CKM}$ taking these values for $\xi_d$ (values of $y_{ij}$ are chosen to fit $m_d$ and $m_s$). 
In addition, the values of $\xi_u$ tend to be smaller than $\xi_d$ due to mass relation $m_b \ll m_t$.

\subsubsection{ $Z'$ interactions with SM fermions} 
The $Z'$ couplings to the SM fermions are written as 
\begin{align}
{\cal L} \supset & -g_X \left( x_\mu \bar \mu \gamma^\mu \mu + x_\tau \bar \tau \gamma^\mu \tau + x_\mu \bar \nu_\mu \gamma^\mu P_L \nu_\mu + x_\tau \bar \nu_\tau \gamma^\mu P_L \nu_\tau +  x_\mu \bar \nu_2  \gamma^\mu P_R\nu_2 + x_\tau \bar \nu_3  \gamma^\mu P_R\nu_3 \right) Z'_\mu \nonumber \\
&+ \frac{g_X}{3} \bar t \gamma^\mu t Z'_\mu +  \frac{g_X}{3} \left( \bar d_\alpha \gamma^\mu P_L d_\beta \Gamma^{d_L}_{\alpha \beta} + \bar d_\alpha \gamma^\mu P_R d_\beta \Gamma^{d_R}_{\alpha \beta} \right) Z'_\mu \ ,
\label{eq:int_Z'}
\end{align}
where $g_X$ is the gauge coupling constant associated with the $U(1)_X$ and the lepton sector 
is given in the flavor basis here. 
The coupling matrices $\Gamma^{d_R}$ and $\Gamma^{d_L}$ for down-type quarks are given approximately by
\begin{equation}
\Gamma^{d_L} \simeq  \left( \begin{array}{ccc} |V_{td}|^2 & V_{ts}V^*_{td} & V_{tb} V^*_{td} \\ V_{td} V^*_{ts} & |V_{ts}|^2 & V_{tb} V^*_{ts} \\ V_{td} V^*_{tb} & V_{ts} V^*_{tb} & |V_{tb}|^2  \end{array} \right), \quad 
\Gamma^{d_R} \simeq  \left( \begin{array}{ccc} 0 & 0 & 0 \\ 0 & 0 & 0 \\ 0 & 0 & 1 \end{array} \right),
\label{eq:ckm}
\end{equation}
where $V_{qq'}$'s are the CKM matrix elements.  We have applied the relation 
$V_{CKM} \simeq D_{ L}$, as we discussed above.
In our model the $Z'$ mass, $m_{Z'}$, is dominantly given by the VEV of SM singlet scalar field as 
discussed below.

At this point, $x_\mu$ is an arbitrary parameter requiring only anomaly cancellation condition Eq.~(\ref{Eq:anomaly}).
This value will be fixed to obtain negative $\Delta C_9^\mu$ and to realize minimal scalar sector.
The mass of $Z'$ can be a free parameter since it is given by new gauge coupling $g_X$ and scalar singlet VEV 
where we have freedom to chose the VEV even if the gauge coupling is fixed.

\subsubsection{Effective interaction for $b \to s \mu^+ \mu^-$}

\begin{figure}[tb]
\begin{center}
\includegraphics[width=75mm]{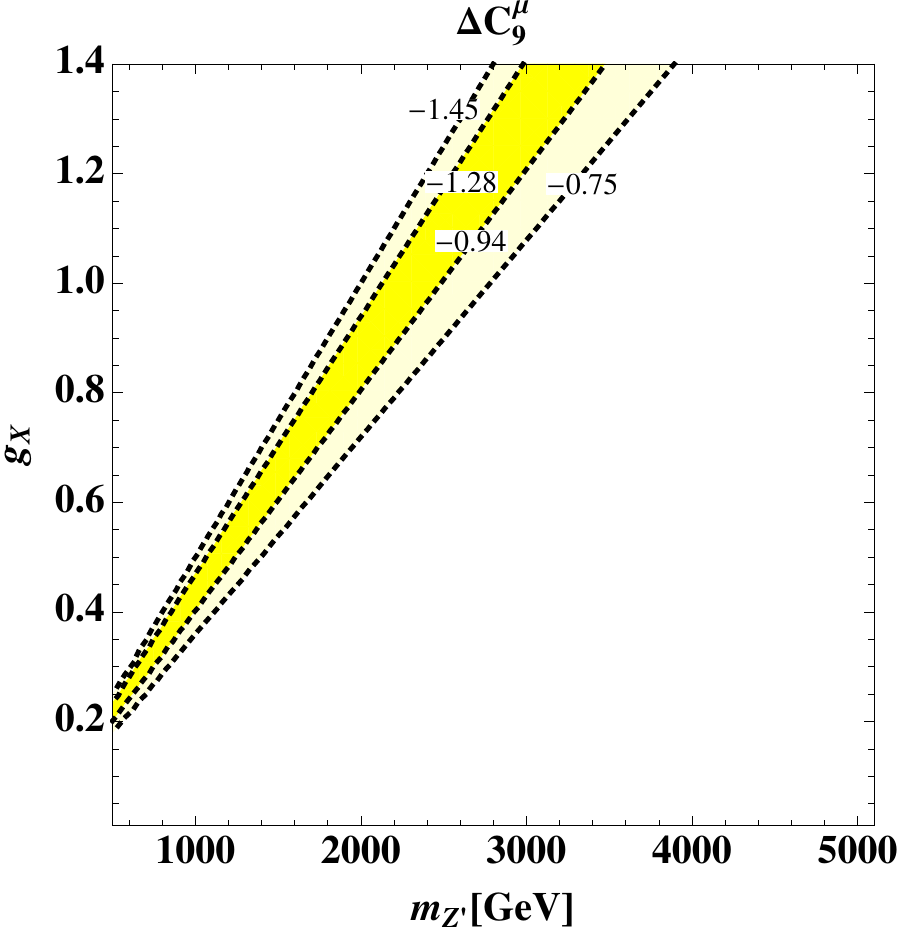} 
\caption{ The contours showing $Z'$ contribution to $\Delta C_9^\mu$ on the $m_{Z'}$-$g_X$ plane with $x_\mu = -\frac{1}{3}$ where yellow(light-yellow) region corresponds to 1$\sigma$ (2$\sigma$) region from global fit in Ref.~\cite{Capdevila:2017bsm}. }
\label{fig:C9}
\end{center}
\end{figure}

Gauge interactions in Eq.~(\ref{eq:int_Z'}) induce the effective Hamiltonian for $b \to s \mu^+ \mu^-$ process such that
\begin{align}
\Delta H_{\rm eff} &= -\frac{x_\mu g_{X}^2 V_{tb} V_{ts}^* }{3 m_{Z'}^2}  (\bar s \gamma^\mu P_L b) (\bar \mu \gamma_\mu \mu) + h.c. \nonumber \\
& = \frac{x_\mu g_X^2}{3 m_{Z'}^2} \left( \frac{\sqrt{2} \pi}{G_F \alpha_{em}} \right) \left( \frac{-  4 G_F}{\sqrt{2} } \frac{\alpha_{em}}{4 \pi} V_{tb} V_{ts}^* \right)(\bar s \gamma^\mu P_L b) (\bar \mu \gamma_\mu \mu) + h.c.,
\end{align}
where $G_F$ is the Fermi constant and $\alpha_{em}$ is the electromagnetic fine structure constant. 
We thus obtain the $Z'$ contribution to Wilson coefficient $\Delta C_9^\mu$ as
\begin{equation}
\Delta C_9^\mu = \frac{x_\mu g_X^2}{3 m_{Z'}^2} \left( \frac{\sqrt{2} \pi}{G_F \alpha_{em}} \right) 
\simeq 2.78 \times x_\mu \left( \frac{ g_X}{0.62} \right)^2 \left( \frac{1.5 \ {\rm TeV}}{m_{Z'}} \right)^2.
\end{equation} 
In order to obtain $\Delta C_9^\mu \sim -1$, $x_\mu$ should be negative and $g_X$ is 
required to be $\sim 0.6$ for $m_{Z'} = 1.5$ TeV and $x_\mu = - \frac{1}{3}$.
Figure~\ref{fig:C9} shows the contour of $\Delta C_9^\mu$ in the $( m_{Z'} , g_X )$ plane where 
we took $x_\mu = - \frac{1}{3}$ where the yellow(light-yellow) region corresponds to 
1$\sigma$ (2$\sigma$) region from global fit in Ref.~\cite{Capdevila:2017bsm}.

\begin{table}[t]
\begin{center} 
\begin{tabular}{|c||c|c|c|c||c|}\hline\hline  
Fields & ~$\Phi_1$~ & ~$\Phi_2$~ & ~$\varphi_1$~ &~$\varphi_2$~ & ~$\chi$~ \\\hline 
 $SU(2)_L$ & $\bf{2}$  & $\bf{2}$  & $\bf{1}$ & $\bf{1}$ & $\bf{1}$    \\\hline 
$U(1)_Y$ & $\frac12$ & $\frac12$  & $0$ & $0$  & $0$     \\\hline
 $U(1)_{X}$ & $-\frac13$ & $0$  & $\frac13$ & $1$  & $\frac56$      \\\hline
\end{tabular}
\caption{Scalar fields and extra fermion $\chi$ in the minimal model and their representation under $SU(2) \times U(1)_Y \times U(1)_X$ where these fields are color singlet.}
\label{tab:2}
\end{center}
\end{table}

\subsection{Minimal model}

Here we consider the minimal cases for choosing $U(1)_X$ charges of leptons as
\begin{equation}
x_\mu = -\frac{1}{3}, \quad x_\tau = \frac{4}{3}.
\end{equation}
In this case we add two $SU(2)_L$ singlet scalar fields:
\begin{equation}
\varphi_1 : ({\bf 1},{\bf 1})(0,1/3), \quad \varphi_2 : ({\bf 1},{\bf 1})(0,1), 
\end{equation}
where $\varphi_1$ is also necessary to induce $\Phi_1^\dagger \Phi_2$ terms
\footnote{Note that we need one more scalar singlet to generate neutrino mass when $x_\mu \neq -1/3$.}, while $\varphi_2$ is added for generating the $23(32)$ element of Majorana mass 
matrix of right-handed neutrino.
 Note that we obtain a massless Goldstone boson from two Higgs doublet sector without $\varphi_1$ due to an additional global symmetry. 
In addition we introduce additional Dirac fermion $\chi$ of mass $m_X$ with $U(1)_X$ charge 
$5/6$, which can be 
our DM candidate since its stability is guaranteed due to fractional charge assignment under 
$U(1)_X$.  Note that  the stability of Dirac fermion DM $\chi$  is guaranteed by remnant $Z_2$ 
symmetry after $U(1)_X$ symmetry breaking: particles with $U(1)_X$ charge $2n/6$ ($n$ is integer) 
are $Z_2$  even and those with $U(1)_X$ charge $(2n+1)/6$ are $Z_2$ odd, 
since $U(1)_X$ symmetry is 
      broken by VEVs of scalar fields $\varphi_1$, $\varphi_2$ and $\Phi_1$ whose charges 
      correspond to $2n/6$~\cite{Krauss:1988zc}.  
We summarize the charge assignment of scalar fields and new fermion in Table~\ref{tab:2}.
In the later analysis, we will adopt this minimal setting.

In our set up, the full scalar potential for scalar fields in our model is given by
\begin{align}
V_{}=& -\mu (\Phi_1^\dagger \Phi_2 \varphi_1^*+{\rm h.c.}) + \mu_{11}^2 |\Phi_1|^2 + \mu_{22}^2 |\Phi_2|^2 
+ \mu_{\varphi_1}^2  |\varphi_1|^2 + \mu_{\varphi_2}^2  |\varphi_2|^2 \nonumber \\
&+ \frac{\lambda_1}{2} |\Phi_1|^4 + \frac{\lambda_2}{2} |\Phi_2|^4 + \lambda_3 |\Phi_1|^2 |\Phi_2|^2 
+ \lambda_4 |\Phi_1^\dagger \Phi_2|^2 + \lambda_{\varphi_1} |\varphi_1|^4 + \lambda_{\varphi_2} |\varphi_2|^4  \nonumber \\
&
+ \lambda_{\Phi_1 \varphi_1} |\Phi_1|^2 |\varphi_1|^2 + \lambda_{\Phi_2 \varphi_1} |\Phi_2|^2 |\varphi_1|^2
+ \lambda_{\Phi_1 \varphi_2} |\Phi_1|^2 |\varphi_2|^2 + \lambda_{\Phi_2 \varphi_2} |\Phi_2|^2 |\varphi_2|^2 + \lambda_{\varphi_1 \varphi_2} |\varphi_1|^2 |\varphi_2|^2 \nonumber \\
&
- \lambda_X (\varphi_1^3 \varphi_2^* + h.c.),
\label{eq:Fullpotential}
\end{align}   
where we assumed all the coupling constants are real for simplicity.
The VEVs of singlet scalar fields are written by $\sqrt{2} \langle \varphi_1 \rangle = v_{\varphi_1}$ and $\sqrt{2} \langle \varphi_2 \rangle = v_{\varphi_2}$.
In our scenario, we assume $v_{\varphi_1}^2 \gg v_{\varphi_2}^2 \gg v_{1,2}^2$ and $U(1)_X$ symmetry is spontaneously broken at a scale higher than the electroweak scale.
We then approximately obtain VEVs of $\varphi_{1,2}$ from the condition $\partial V/\partial v_{\varphi_{1,2}} =0$:
\begin{equation}
v_{\varphi_1} \simeq \sqrt{\frac{- \mu_{\varphi_1}^2}{\lambda_{\varphi_1}}}, \quad v_{\varphi_2} \simeq \frac{ \lambda_X  v_{\varphi_1}^3}{4 \mu_{\varphi_2}^2 + 2 \lambda_{\varphi_1 \varphi_2} v_{\varphi_1}^2},
\end{equation}
where the above assumption for VEV hierarchy can be consistent requiring $\lambda_X v_{\varphi_1}^2 \ll \mu_{\varphi_2}^2$.
Then  the mass of the $Z'$ boson is approximately given by
\begin{equation}
m_{Z'} \simeq \frac{1}{3} g_X v_{\varphi_1}.
\end{equation}
Then a typical value of the $\varphi_1$ VEV is $v_{\varphi_1} \simeq 7.5 \times (m_{Z'}/1.5 \ {\rm TeV})(0.6/g_X) \ {\rm TeV}$ in our scenario.
Note that the $Z$--$Z'$ mass mixing is highly suppressed by $v_1^2/v_{\varphi_1}^2$ factor which is $\sim  10^{-5}$ for $\tan \beta = v_2/v_1 = 10$ and $v_{\varphi_1} = 7. 5$ TeV.
Thus we will ignore this effect in our analysis~\footnote{The $Z$--$Z'$ mixing effect is constrained by precision measurements of $Z \bar f_{SM} f_{SM}$ coupling at the LEP experiments where the upper bound of the mixing $\theta_{ZZ'}$ is around $\sim 10^{-3}-10^{-4}$~\cite{Tanabashi:2018oca,Langacker:2008yv}. Thus our mixing angle is sufficiently smaller than the bound.  }.

After $U(1)_X$ symmetry breaking, we obtain two-Higgs doublet potential effectively~\footnote{Here we do not consider scalar bosons from $\varphi_{1,2}$ since they are assumed to be much heavier than those from Higgs doublets and mixing among singlet and doublet scalars will be small. }:
\begin{align}
V_{THDM} = & m_1^2 |\Phi_1|^2 + m_2^2 |\Phi_2|^2 - (m_3^2 \Phi_1^\dagger \Phi_2 + h.c. ) \nonumber \\
& + \frac{\lambda_1}{2} |\Phi_1|^4 + \frac{\lambda_2}{2} |\Phi_2|^4 + \lambda_3 |\Phi_1|^2 |\Phi_2|^2 + \lambda_4 |\Phi_1^\dagger \Phi_2|^2, \\
m_{1(2)}^2 = & \mu_{11(22)}^2 + \frac{1}{2} \lambda_{\Phi_{1(2)} \varphi_1} v_{\varphi_1}^2 + \frac{1}{2} \lambda_{\Phi_{1(2)} \varphi_2} v_{\varphi_2}^2, \quad
m_3^2 = \frac{1}{\sqrt{2}} \mu v_{\varphi_1}. 
\end{align}
Here we write $\Phi_{i}$ ($i=1,2$) as 
\begin{equation}
\Phi_i = \begin{pmatrix} w_i^+ \\ \frac{1}{\sqrt{2}} (v_i + h_i + i z_i) \end{pmatrix}.
\end{equation}
As in the two-Higgs-doublet model (THDM), we obtain mass eigenstate $\{H, h, A, H^\pm \}$ in the two Higgs doublet sector:
\begin{align}
\begin{pmatrix} z_1 (w^+_1) \\ z_2 (w^+_2) \end{pmatrix}  & = \begin{pmatrix} \cos \beta & - \sin \beta \\ \sin \beta & \cos \beta \end{pmatrix} \begin{pmatrix} G_Z (G^+) \\ A(H^+) \end{pmatrix}, \\
\begin{pmatrix} h_1 \\ h_2 \end{pmatrix}  & = \begin{pmatrix} \cos \alpha & - \sin \alpha \\ \sin \alpha & \cos \alpha \end{pmatrix} \begin{pmatrix} H \\ h \end{pmatrix},
\end{align}
where $\tan \beta = v_2/v_1$, $G_Z (G^+)$  is a Nambu-Goldstone boson (NG) absorbed by the $Z(W^+)$ boson, and $h$ is the SM-like Higgs boson.
The masses of $H^\pm$ and $A$ are given as in THDM:
\begin{equation}
m^2_{H^\pm} = \frac{m_3^2}{\sin \beta \cos \beta} - \frac{v^2}{2} \lambda_4, \quad m_A^2 = \frac{m_3^2}{\sin \beta \cos \beta}.
\end{equation}
Mass eigenvalues of CP-even scalar bosons are also obtained by
\begin{align}
& m^2_{H,h} = \frac{1}{2} \left( M_1^2 + M_2^2 \pm \sqrt{(M_1^2 - M_2^2)^2 + 4 M_{12}^4} \right), \\
& M_1^2 = v^2 (\lambda_1 \cos^4 \beta + \lambda_2 \sin^4 \beta) + \frac{v^2}{2} \bar \lambda \sin^2 2 \beta, \\
& M_2^2 = \frac{m_3^2}{\sin \beta \cos \beta} + v^2 \sin^2 \beta \cos^2 \beta (\lambda_1 + \lambda_2 - 2 \bar \lambda), \\
& M^2_{12} = \frac{v^2}{2} \sin 2 \beta (- \lambda_1 \cos^2 \beta + \lambda_2 \sin^2 \beta) + \frac{v^2}{2} \bar \lambda \sin 2 \beta \cos 2 \beta ,
\end{align}
where $\bar \lambda = \lambda_3 + \lambda_4$ and lighter mass eigenvalue $m_h$ is identified as the SM-like Higgs mass.

Note that Higgs bosons in doublet interact with $Z'$ and three point couplings can be obtained such that
\begin{align}
(D_\mu H_1)^\dagger (D^\mu H_1) \supset & i \frac{g_X}{3} Z'^\mu (w^+_1 \partial_\mu w^-_1  - w_1^- \partial_\mu w_1^+ ) 
+ \frac{2 g_X}{3} Z'^\mu (h_1 \partial_\mu z_1  - z_1 \partial_\mu h1  ) \nonumber \\
\supset & i \frac{g_X \sin^2 \beta}{3} Z'^\mu (H^+ \partial_\mu H^- - H^- \partial_\mu H^+)
+ \frac{2 g_X \sin \beta \sin \alpha}{3} Z'^\mu (h \partial_\mu A - A \partial_\mu h) \nonumber \\
& + \frac{2 g_X \sin \beta \cos \alpha}{3} Z'^\mu (A \partial_\mu H - H \partial_\mu A).
\end{align}
Thus $Z'$ can decay into $HA$, $hA$ and $H^+H^-$ pair.

Here we briefly comment on deviation in the couplings of the SM-like Higgs $h$ and constraint in the scalar sector in the model. 
The Yukawa interactions with $h$ are given by Eq.~(\ref{Eq:YukawaH}) in the Appendix.
In particular, we have flavor violating interaction associated with $\xi^{u,d}$ coupling.
In our analysis, we assume the interactions are SM-like that can be realized taking large $\tan \beta$ and alignment limit of $\cos (\alpha  - \beta) \simeq 0$. 
Note also that new scalar bosons do not contribute to explanation of $b \to s \mu^{+} \mu^{-}$ anomalies in our scenario except for 
relaxing the constraint from $B_s$--$\bar B_s$ mixing as we discuss below; we can fit the data with the mass value of $\sim 500$ to $\sim 1000$ GeV for exotic scalar bosons from two-Higgs doublet sector.
In such a mass region, we can find a parameter to avoid collider constraints for exotic scalar production like that of charged scalar bosons~\cite{Aaboud:2018cwk}.
We thus just assume new scalar bosons are sufficiently heavy and we can avoid constraints from scalar boson search at the LHC.
Discussion of scalar sector can be referred to, for example, Refs.~\cite{Crivellin:2015lwa, Bian:2017xzg}.

\section{Neutrino mass and flavor constraints}

In this section we formulate neutrino mass matrices (both Dirac and Majorana mass matrices), 
and explore constraints from flavor physics such as $\mu \to e \gamma$, $\mu \to e$ conversion 
and  $B_s$--$\bar B_s$ mixing.

\subsection{Neutrino mass matrices}

The Yukawa interactions for leptons are given by
\begin{align}
- \mathcal{L} \ \supset \ & y^e_{aa} \bar L_{a L} e_{a R} \Phi_2 + y^\nu_{aa} \bar L_{a L} \nu_{a R} \tilde \Phi_2 + \tilde y^e_{12} \bar L_{1 L} \mu_{ R} \Phi_1  + \tilde y^\nu_{21} \bar L_{2 L} \nu_{1 R} \tilde \Phi_1 \nonumber \\
& + M \bar \nu_{1R}^c \nu_{1 R} +   Y_{12} \bar \nu_{1R}^{ c}  \nu_{2 R} \varphi_1^*  + Y_{23} \bar \nu_{2 R}^{ c} \nu_{3 R} \varphi_2^* + h.c.,
\end{align}
where $a=1,2,3$ and $Y_{ab} = Y_{ba}$. After the symmetry breaking, Dirac and Majorana mass matrices for neutrinos have the structure of
\begin{equation}
M_D = \begin{pmatrix} (M_D)_{11} & 0 & 0 \\ (M_D)_{21} & (M_D)_{22} & 0 \\ 0 & 0 & (M_D)_{33} \end{pmatrix},  \quad
M_{\nu_R} = \begin{pmatrix} (M_{\nu_R})_{11} & (M_{\nu_R})_{12} & 0 \\ (M_{\nu_R})_{21} & 0 & (M_{\nu_R})_{23} \\ 0 & (M_{\nu_R})_{32} & 0 \end{pmatrix},
\end{equation}
where the elements of the mass matrices are given by
\begin{align}
& (M_{D})_{aa} = \frac{1}{\sqrt{2}} y^\nu_{aa} v_2, \quad (M_D)_{21} = \frac{1}{\sqrt{2}} \tilde y_{21} v_1, \nonumber \\
& (M_{\nu_R})_{11} = M, \quad (M_{\nu_R})_{12(21)} = \frac{1}{\sqrt{2}} Y_{12} v_{\varphi_1}, \quad (M_{\nu_R})_{23(32)} = \frac{1}{\sqrt{2}} Y_{23} v_{\varphi_2}.
\end{align} 
The active neutrino mass matrix is given by type-I seesaw mechanism:
\begin{align}
& m_\nu  \simeq - M_D M_{\nu_R}^{-1} M_D^T \nonumber \\
&= \begin{pmatrix} \frac{(M_D)_{11}^2}{(M_{\nu_R})_{11}} & \frac{(M_D)_{11} (M_D)_{21}}{(M_{\nu_R})_{11}} & -\frac{(M_D)_{11} (M_D)_{33} (M_{\nu_R})_{12}}{(M_{\nu_R})_{11} (M_{\nu_R})_{32}} \\ 
\frac{(M_D)_{11} (M_D)_{21}}{(M_{\nu_R})_{11}} & \frac{(M_D)_{21}^2}{(M_{\nu_R})_{11}} &  \frac{(M_D)_{33} (M_D)_{22}}{(M_{\nu_R})_{32}} \left( 1 - \frac{(M_D)_{21}(M_{\nu_R})_{12}}{(M_{\nu_R})_{11} (M_D)_{22}} \right) \\  
-\frac{(M_D)_{11} (M_D)_{33} (M_{\nu_R})_{12}}{(M_{\nu_R})_{11} (M_{\nu_R})_{32}} & \frac{(M_D)_{33} (M_D)_{22}}{(M_{\nu_R})_{32}} \left( 1 - \frac{(M_D)_{21}(M_{\nu_R})_{12}}{(M_{\nu_R})_{11} (M_D)_{22}} \right) & \frac{ (M_D)_{33}^2 (M_{\nu_R})_{12}^2}{(M_{\nu_R})_{11} (M_{\nu_R})_{23}^2}
\end{pmatrix}.
\end{align}
Note that our neutrino mass matrix does not have zero structure and neutrino oscillation data 
can be easily fit.   Here we  do not carry out  further analysis of the neutrino phenomenology in this paper.

\subsection{Charged lepton mass matrices}

The charged lepton mass matrix is given by
\begin{align}
M^e &= \frac{1}{\sqrt{2}} \begin{pmatrix} y^e_{11} v_2 & \tilde y^e_{12} v_1 & 0 \\  0 &  y^e_{22} v_2 & 0 \\ 0 & 0 &  y^e_{33} v_2 \end{pmatrix}  \equiv \begin{pmatrix} m^e_{11}  & \delta m^e_{12} & 0 \\  0 &  m^e_{22} & 0 \\ 0 & 0 &  m^e_{33}  \end{pmatrix}.
\end{align}
For $\delta m^e_{12} \ll m^e_{22}$, the mass matrix can be diagonalized in good approximation as
\begin{align}
& \begin{pmatrix} m_e & 0 & 0 \\ 0 & m_\mu & 0 \\ 0 & 0 & m_\tau \end{pmatrix} \simeq V_L^e M^e (V_R^e)^\dagger, \\
& V_R^e \simeq {\bm 1}, \quad V_L^e \simeq \begin{pmatrix} 1 & - \epsilon & 0 \\ \epsilon & 1 & 0 \\ 0 & 0 & 1 \end{pmatrix},
\end{align}
where $\epsilon = \delta m^e_{12}/m^e_{22}$ we also find $m_e \simeq m^e_{11}$, $m_\mu \simeq m^e_{22}$ and $m^e_{33} = m_\tau$.

\subsection{Charged lepton flavor violation}

 \begin{figure}[tb]
\begin{center}
\includegraphics[width=90mm]{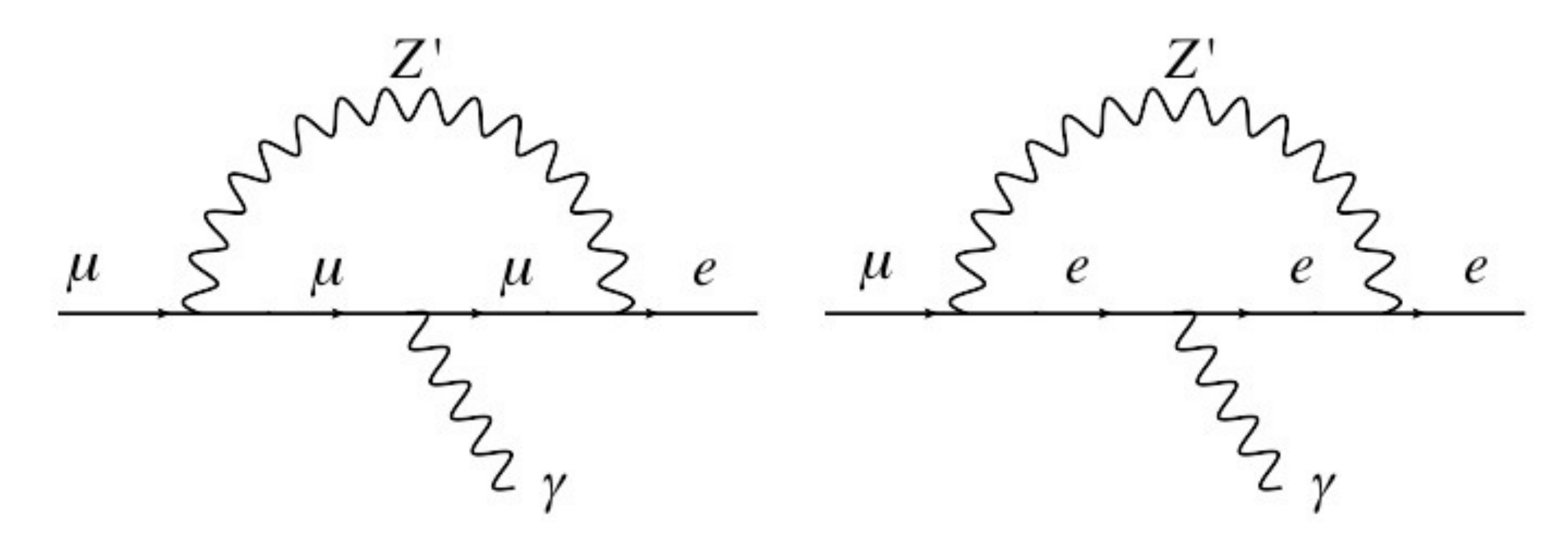} 
\caption{ One loop diagrams inducing $\mu \to e \gamma$ process.}
\label{fig:diagram}
\end{center}
\end{figure}

\begin{figure}[tb]
\begin{center}
\includegraphics[width=75mm]{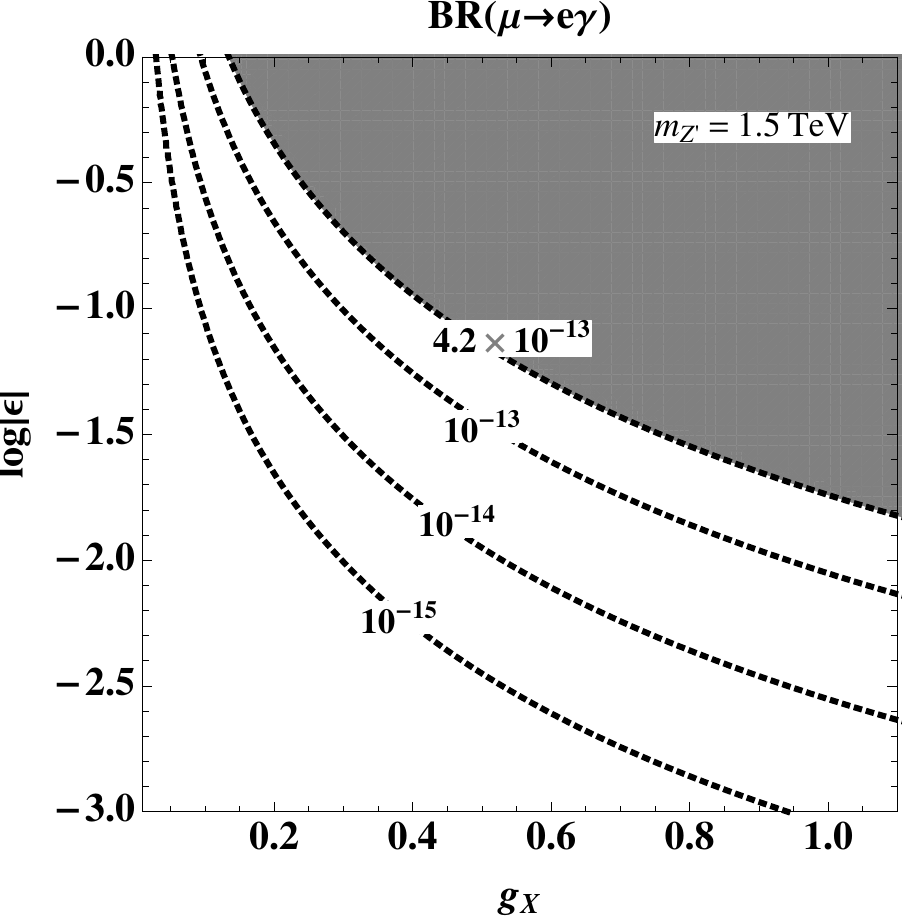} 
\includegraphics[width=75mm]{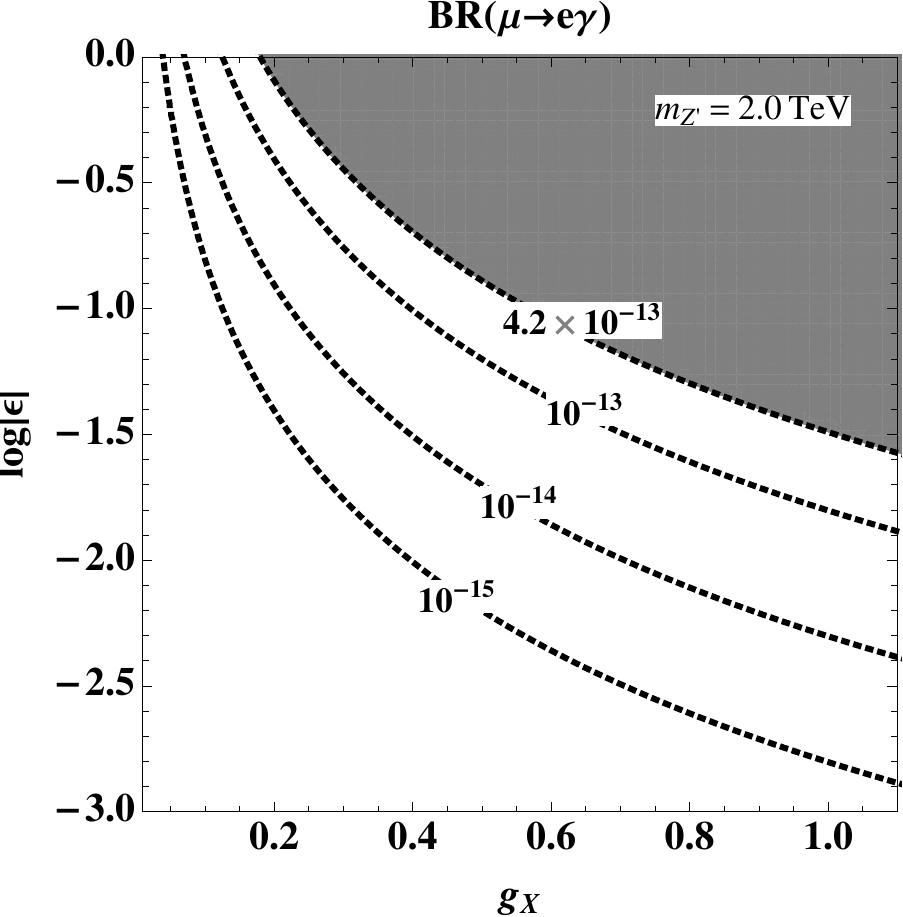} 
\caption{ $BR(\mu \to e \gamma)$ as a function of $\{g_X, \log |\epsilon| \}$ fixing $m_{Z'} = 1.5(2.0)$ TeV for left(right) plot where the shaded regions are excluded.}
\label{fig:LFV}
\end{center}
\end{figure}

Here we  consider charged lepton flavor violation (cLFV) in the model associated with $Z'$. 
The $Z'$ gauge interactions for mass eigenstates of charged leptons are given by
\begin{align}
 {\cal L} \supset - \frac{g_X}{3} \bar \ell_i \gamma^\mu \left[ V_L^e \begin{pmatrix} 0 & 0 &0 \\ 0 & -1 & 0 \\ 0 & 0 & 4 \end{pmatrix}  V_L^{e\dagger} \right]_{ij} P_L \ell_j Z'_\mu - \frac{g_X}{3} \bar \ell_i \gamma^\mu \begin{pmatrix} 0 & 0 &0 \\ 0 & -1 & 0 \\ 0 & 0 & 4 \end{pmatrix}_{ij} P_R \ell_j Z'_\mu, 
\end{align}
where the flavor violating structure for left-handed charged lepton currents is given by
\begin{align}
V_L^e \begin{pmatrix} 0 & 0 &0 \\ 0 & -1 & 0 \\  0 & 0 & 4 \end{pmatrix}  V_L^{e\dagger} \simeq 
\begin{pmatrix} - \epsilon^2 & \epsilon & 0 \\ \epsilon & -1 & 0 \\ 0 & 0 & 4 \end{pmatrix}.
\end{align}
Thus we have LFV interaction for $e$ and $\mu$.
Then we first consider $\mu \to e \gamma$ process induced by $Z'$ loop in Fig.~\ref{fig:diagram} where the left diagram gives dominant contribution due to suppression by $\epsilon$.
Estimating the loop diagram we obtain dominant contribution to the decay width for the 
$\mu \to e \gamma$ process such that
\begin{align}
& \Gamma_{\mu \to e \gamma} \simeq \frac{e^2 m_\mu^3}{16 \pi}  |a_R|^2 , \\
& a_R \simeq \frac{e \epsilon g_X^2 m_\mu}{144 \pi^2} \int_0^1 dx dy dz \delta(1-x-y-z) \frac{2x (1+y) }{[(x^2-x) + xz + y+ z] m_\mu^2 + x  m_{Z'}^2 }.
\end{align}
Branching ratio for the LFV process is given by 
\begin{equation}
BR(\mu \to e \gamma) = \frac{\Gamma_{\mu \to e \gamma}}{\Gamma_{\mu \to e \bar \nu_e \nu_\mu}} \simeq \frac{12 \alpha}{G_F^2 m_\mu^2} |a_R|^2,
\end{equation}
where $G_F \simeq 1.17 \times 10^{-5} \ {\rm GeV}^{-2}$ is the Fermi constant and $\alpha \simeq 1/137$ is the fine structure constant.
In Fig.~\ref{fig:LFV}, we show $BR(\mu \to e \gamma)$ on $\{g_X, \log |\epsilon| \}$ plane fixing $m_{Z'} = 1.5 (2.0)$ TeV where the shaded regions are excluded by the current constraint
$BR (\mu \to e \gamma) \lesssim 4.2 \times 10^{-13}$ by the MEG experiment~\cite{TheMEG:2016wtm}.
Further parameter region will be explored in future with improved sensitivity~\cite{Baldini:2018nnn}.

\begin{figure}[tb]
\begin{center}
\includegraphics[width=75mm]{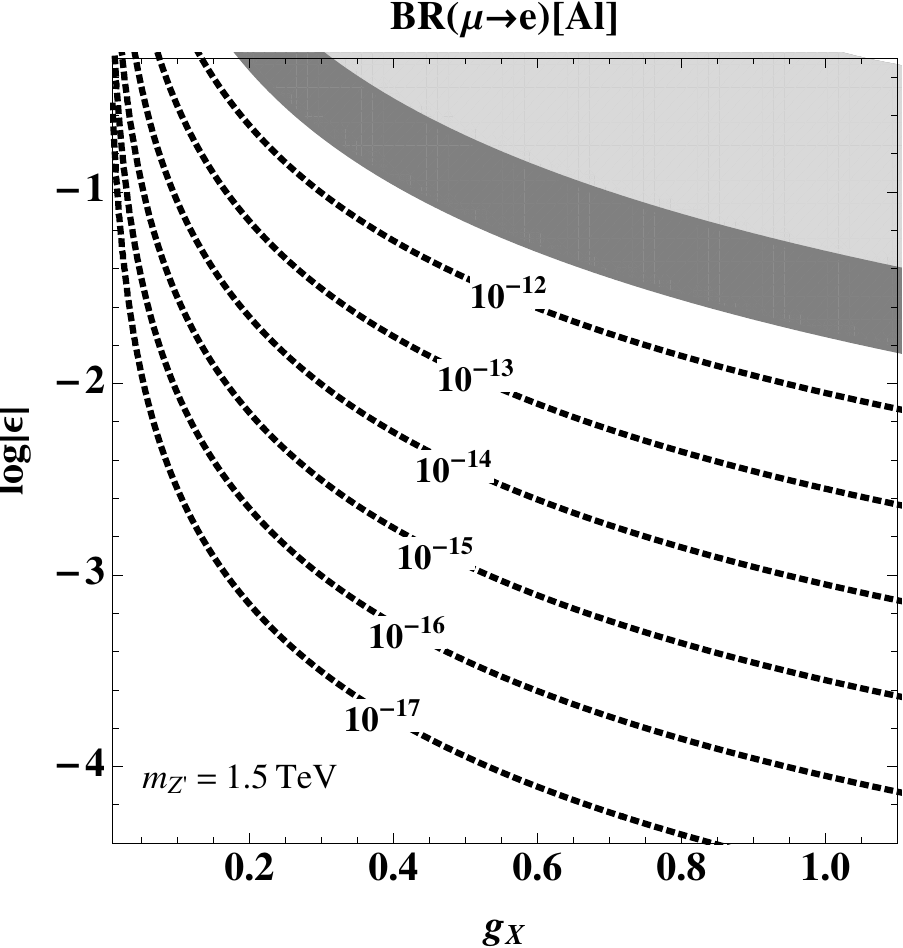} 
\includegraphics[width=75mm]{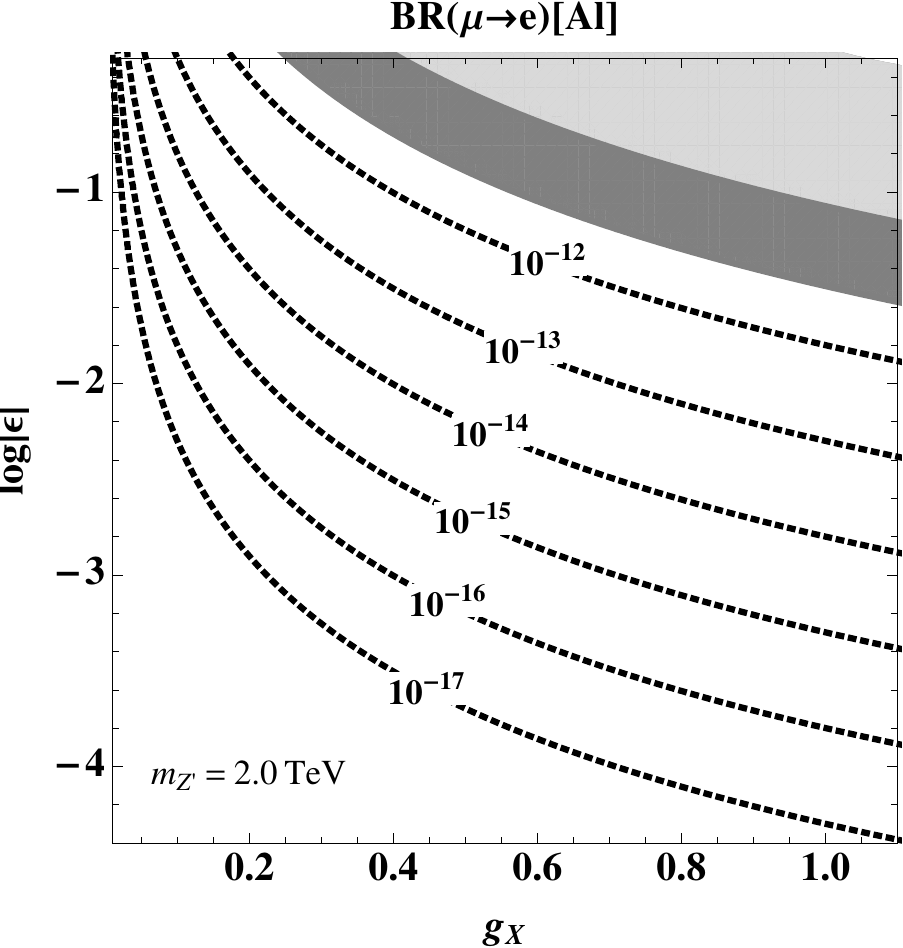} 
\caption{ $BR(\mu \to e)$ on $^{27}_{13}$Al as a function of $\{g_X, \log |\epsilon| \}$ fixing $m_{Z'} = 1.5(2.0)$ TeV for left(right) plot where gray(light-gray) shaded region is excluded by current $\mu \to e \gamma$ BR ($\mu \to e$ BR on $^{197}_{79}$Au~\cite{Bertl:2006up}) constraints.}
\label{fig:MtoE}
\end{center}
\end{figure}

Here we also discuss $\mu \to e$ conversion via $Z'$ exchange. 
In our case, the relevant effective Lagrangian for the process is derived 
as follows~\cite{Kuno:1999jp, Kitano:2002mt, Davidson:2018kud}
\begin{equation}
\mathcal{L}_{eff} = - \frac{4 G_F}{\sqrt{2}} \sum_{N=p,n} \left[ C_{VL}^{NN} \bar e \gamma^\alpha P_L \mu \bar N \gamma_\alpha N 
+ C_{AL}^{NN} \bar e \gamma^\alpha P_L \mu \bar N \gamma_\alpha \gamma_5 N  \right],
\end{equation}
where the corresponding coefficients are given by
\begin{equation}
C_{VL}^{pp(nn)} = -C_{AL}^{pp(nn)}  = (2) \frac{\sqrt{2} \epsilon g_X^2 |V_{td}|^2 }{216 G_F m_{Z'}^2}. 
\end{equation}
Then we obtain the  spin-independent contribution to the BR for $\mu \to e$ conversion on a nucleus such that
\begin{equation}
BR(\mu \to e) = \frac{32 G_F^2 m_\mu^5}{\Gamma_{cap}} \left| C_{VL}^{pp} V^{(p)} + C_{VL}^{nn} V^{(n)} \right|^2,
\end{equation}
where $\Gamma_{cap}$ is the rate for the muon to transform to a neutrino by capture on the nucleus, and $V^{(p,n)}$ is the integral over the nucleus for lepton wavefunctions with corresponding nucleon density.
The values of $\Gamma_{cap}$ and $V^{(n,p)}$ depend on target nucleus and those for $^{197}_{79}$Au and $^{27}_{13}$Al are given in Table.~\ref{tab:mue-conv}~\cite{Kitano:2002mt, Suzuki:1987jf}.
In Fig.~\ref{fig:MtoE}, we show $BR(\mu \to e)$ for $^{27}_{13}$Al on $\{g_X, \log |\epsilon| \}$ plane fixing $m_{Z'} = 1.5(2.0)$ TeV in left(right)-panel where gray(light-gray) shaded region is excluded by current $\mu \to e \gamma$ BR ($\mu \to e$ BR on $^{197}_{79}$Au~\cite{Bertl:2006up}) constraints.
We find that large parameter region can be explored by $\mu \to e$ conversion measurement since its sensitivity will reach $\sim 10^{-16}$ on $^{27}_{13}$Al nucleus in future experiments~\cite{Kuno:2013mha, Carey:2008zz}.

\begin{table}[t]
\begin{center}
\begin{tabular}{|c|c|c|c|} \hline
Nucleus $^A_Z N$ & $V^{p}$ & $V^{n}$ & $ \Gamma_{\rm capt}~[10^6{\rm sec}^{-1}]$ \\ \hline
$^{27}_{13}$Al & $0.0161$ & $0.0173$ & $0.7054$   \\
$^{197}_{79}$Au & $0.0974$ & $0.146$ & $13.07$    \\ \hline
\end{tabular}
\caption{
A summary of parameters for the $\mu-e$ conversion formula for $^{27}_{13}$Al and $^{197}_{79}$Au nuclei~\cite{Kitano:2002mt, Suzuki:1987jf}.}
\label{tab:mue-conv}
\end{center}
\end{table}

We next consider  the LFV B decay $B_s \to \mu^\pm e^\mp$ which is related to $C_9^\mu$ above.
It is because that the process is induced from $C_{10}^{\mu e}$ which is obtained as 
$C_{10}^{\mu e} = - \epsilon \Delta C_9^\mu$ in the model.
The branching ratio can be given by
\begin{equation}
BR(B_s \to \mu e) = \left| \frac{C_{10}^{\mu e}}{C_{10}^{SM}} \right|^2 BR(B_s \to \mu^+ \mu^-)_{SM} \simeq |0.24 \times \epsilon \Delta C_9^\mu|^2 BR(B_s \to \mu^+ \mu^-)_{SM}, 
\end{equation}
where we used $C_{10}^{SM} (\mu_b) \simeq -4.2$ and $BR(B_s \to \mu^+ \mu^-)_{SM} = (3.65 \pm 0.23) \times 10^{-9}$ is the SM prediction for  the BR of $B_s \to \mu^+ \mu^-$.
We find that $BR(B_s \to \mu e) < 10^{-11}$ in the parameter region satisfying the constraint from $BR(\mu \to e \gamma)$ which is well below the current constraint.

 Here we also discuss the branching ratio for $B \to K^{(*)} \mu e$ through lepton flavor violating $Z'$ coupling. It is suppressed compared to $BR (B \to K^{(*)} \mu \mu)$ by a factor of $|\epsilon \Delta C_9^{\mu}/C_9^\mu|^2 \sim 10^{-3}$ for $\Delta C_9^\mu = -1$ and $\epsilon = 0.1$. 
Thus the BR is small as order of $10^{-10} - 10^{-9}$ and it is well below current bound and challenging to search for the signal at the future experiments such as (upgraded) LHCb~\cite{Bediaga:2012uyd} and Belle II~\cite{Kou:2018nap}.


\subsection{Constraint from neutrino trident process and $Z'$ contribution to muon $g-2$}
$U(1)_X$ gauge coupling and $Z'$ mass are constrained by the neutrino trident process $\nu N \to \nu N \mu^+ \mu^- $ where $N$ is a nucleon~\cite{Altmannshofer:2014pba}. 
The bound is approximately given by $m_{Z'}/g_X \gtrsim 550 \ {\rm GeV}$ for $m_{Z'} > 1$ GeV.
We then consider parameter region of $\{m_{Z'}, g_X\}$ satisfying this bound.

The observed muon magnetic dipole moment is deviated from the SM prediction as $\Delta a_\mu = (26.1 \pm 8.0) \times 10^{-10}$~{\cite{bennett}} (muon $g-2$). 
The $Z'$ boson can contribute to muon $g-2$ at one loop level as
\begin{align}
\Delta a_{\mu}^{Z'}\approx \frac{g_{X}^2 x_\mu^2}{4 \pi^2}\int_0^1 da \frac{r a (1-a)^2}{r(1-a)^2+a}, 
\end{align}
where $r\equiv(m_\mu/M_{Z'})^2$.
We find that the $Z'$ contribution is small for the parameter region providing $\Delta C_9 \sim -1$; for example $\Delta a_{\mu}^{Z'} \sim 1.7 \times 10^{-12}$ with $m_{Z'} = 1500$ GeV and $g_X = 0.6$.

\subsection{Constraint from $B_s$--$\bar B_s$ mixing }

In our model, $Z'$ and neutral scalar bosons induce flavor changing neutral current (FCNC) 
interactions.   Here we consider constraints  from $B_s$--$\bar B_s$ mixing where other 
$\Delta F =2$ processes are more suppressed by CKM factors.

\begin{figure}[tb]
\begin{center}
\includegraphics[width=75mm]{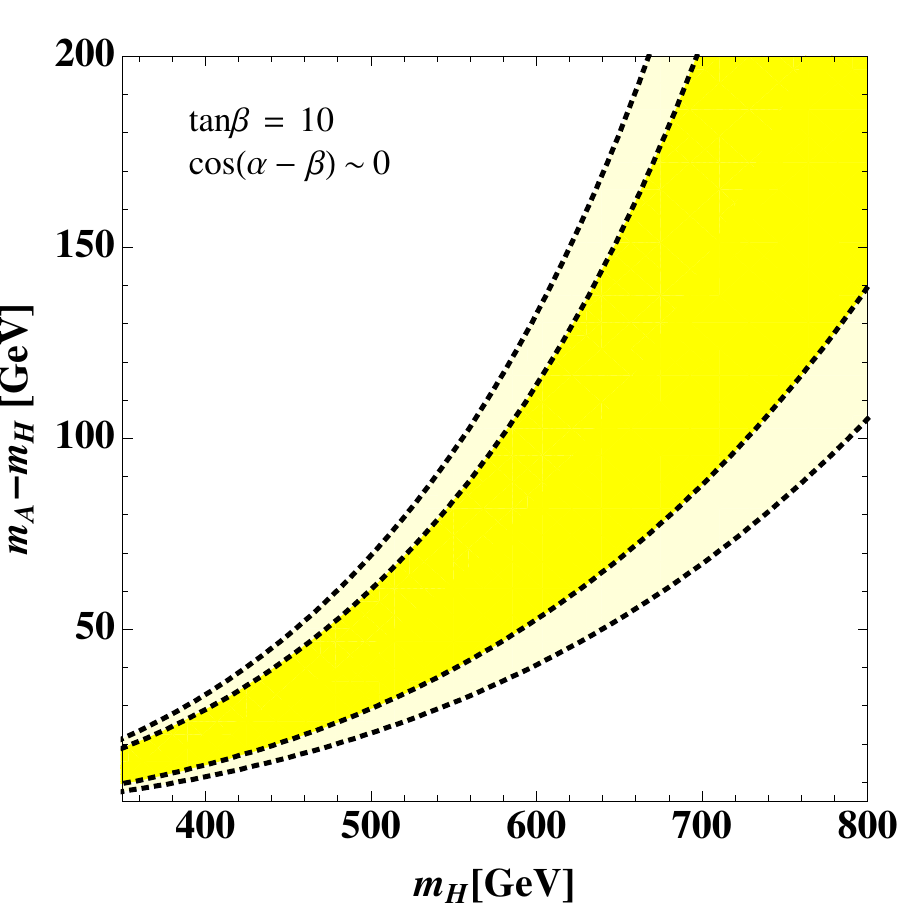} 
\caption{ The allowed region on $\{m_H, m_A -m_H \}$ plane satisfying  $B_s$--$\bar B_s$ 
constraints with fitting $C_9$ to explain $b \to  s \ell^+ \ell^-$ anomalies where the yellow(light yellow) region corresponds to that in Fig.~\ref{fig:C9}. Here we take $\tan \beta = 10$ and $\cos (\alpha - \beta) \sim 0$ as reference values.}
\label{fig:BsBsbar}
\end{center}
\end{figure}

The effective Hamiltonian for the $B_s$--$\bar B_s$ mixing  is given by
\begin{equation}
H_{eff} = C_1 (\bar s \gamma^\mu P_L b) (\bar s \gamma_\mu P_L b) + 
C'_2 (\bar s P_R b)(\bar s P_R b) .
\end{equation}
The relevant Wilson coefficients are 
\begin{equation}
C_1 = \frac12 \frac{g_X^2}{9 m_{Z'}^2} (\Gamma^{d_L}_{sb})^2, \quad C'_2 = \sum_{\eta = h,H,A} \frac{-1}{2 m^2_\eta} (\Gamma^\eta_{sb})^2,
\end{equation}
where $\Gamma^\eta_{qq'}$ is couplings for $\eta \bar q q'$ interactions ($\eta = h, H, A$), the 
explicit expressions of which are given in the Appendix.
Using these Wilson coefficients we obtain ratio between $\Delta m_{B_s}$ in our model and the SM prediction $\Delta m_{B_s}^{SM}$, under large $\tan \beta$ and small $\alpha$, such that
\begin{align}
R_{B_s}  & = \frac{\Delta m_{B_s}}{\Delta m_{B_s}^{SM}} \nonumber \\
& \simeq \frac{g_X^2 (V_{tb} V_{ts}^*)^2}{9 m^2_{Z'}} (8.2 \times 10^{-5} \ {\rm TeV^{-2}})^{-1} \nonumber \\
&+ \left[ 0.12 \cos^2 (\alpha - \beta) \tan^2 \beta + 0.19 \tan^2 \beta \left( \frac{(200 \ {\rm GeV})^2}{m_H^2} - \frac{(200 \ {\rm GeV})^2}{m_A^2} \right) \right],
\end{align}
where the first and second terms in the right-hand side corresponds to contributions from $Z'$ 
and scalars,  respectively~\cite{Crivellin:2015lwa, Arnan:2016cpy, Bazavov:2016nty}.
The allowed range of $R_{B_s}$ is estimated by~\cite{Arnan:2016cpy, Bazavov:2016nty}
\begin{equation}
0.83 < R_{B_s} < 0.99.
\end{equation}
We find that $R_{B_s}$ will be deviated from the allowed range by $Z'$ contribution when  
$\Delta C_9^{\mu} \simeq -1$ is required.
Thus cancellation between $Z'$ and scalar contribution is necessary to satisfy the experimental constraint\footnote{
Similar phenomena  were also observed in the flavor gauge model where $U(1)^{'}$ gauge 
interaction couples only to the right-handed top quark in the interaction basis in the context of the top forward-backward asymmetry and the same sign top pair productions at hadron colliders 
\cite{Ko:2011vd,Ko:2011di,Ko:2012ud,Ko:2012sv}.  
In that model, cancellation between the amplitudes with $t$-channel exchanges of vector and (pseudo)scalar bosons occur in the same sign top pair production through  
$u_R u_R \rightarrow t_R t_R$, which saves the $U(1)$ flavor model from the stringent constraints
from the same sign top pair production at the Tevatron and the LHC.}.
Here we derive allowed parameter region on $\{m_H, m_A -m_H \}$ plane satisfying 
$B_s$--$\bar B_s$  constraints when we fit  $C_9^\mu$ to explain 
$b \to  s \ell^+ \ell^-$ anomalies 
choosing $\tan \beta = 10$ and $\cos (\alpha - \beta) \sim 0$ as reference values.
In Fig.~\ref{fig:BsBsbar}, we show the allowed parameter region where the yellow(light yellow) region corresponds to that in Fig.~\ref{fig:C9}.

\subsection{Prediction on $B\rightarrow K^{(*)} \tau^+ \tau^-$}

Here we discuss $B\rightarrow K^{(*)} \tau^+ \tau^-$ process in our model. 
The branching ratios are given by Wilson coefficient $C_9$ associated with $\tau$ such that~\cite{Capdevila:2017iqn}
\begin{align}
10^7 \times BR(B \to K \tau^+ \tau^-)^{[15, 22]} = & (1.20 + 0.15 \Delta C_9^{\tau} + 0.02 (\Delta C_9^{\tau})^2 ) \nonumber \\
& \pm (0.12 + 0.02 \Delta C_9^{\tau} ), \\
10^7 \times BR(B \to K^* \tau^+ \tau^-)^{[15, 19]} = & (0.98 + 0.38 \Delta C_9^{\tau} + 0.05 (\Delta C_9^{\tau})^2 ) \nonumber \\
& \pm (0.09 + 0.03 \Delta C_9^{\tau} + 0.01 (\Delta C_9^{\tau})^2 ), 
\end{align}
where the superscript indicates the $q^2$ range for the dilepton invariant mass in unit of [GeV$^2$].
For the $b \rightarrow s \tau^+ \tau^-$ channel, we obtain $\Delta C_9^\tau = - 4 C_9^\mu$ from our 
charge assignments, and the BRs are slightly enhanced from the SM prediction by factor $\sim 1.5$. 
However current upper bounds of the BRs are much larger than the prediction as  
$BR(B \to K \tau^+ \tau^-) < 2.25 \times 10^{-3} $~\cite{TheBaBar:2016xwe}.
Therefore it is difficult to test the enhancement effect.

\section{Collider physics and dark matter}

In this section we explore collider physics focusing on $Z'$ production at the LHC and estimate relic density of our DM candidate 
searching for parameter region providing observed value.

\subsection{$Z'$ production at the LHC}
\begin{figure}[tb]
\begin{center}
\includegraphics[width=75mm]{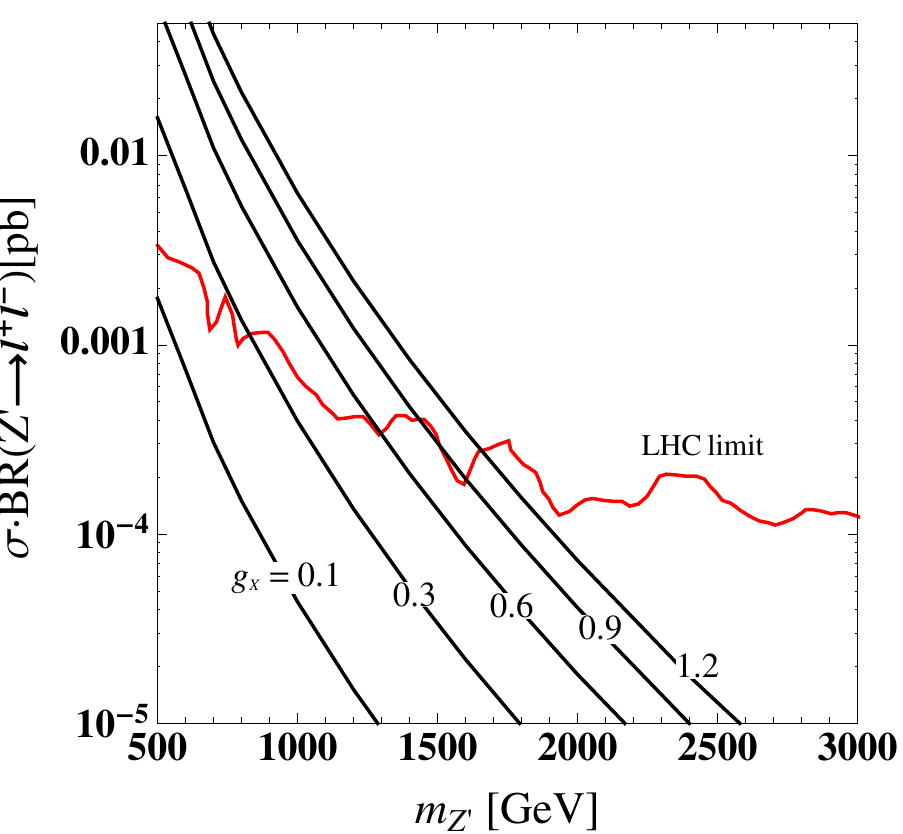} 
\includegraphics[width=75mm]{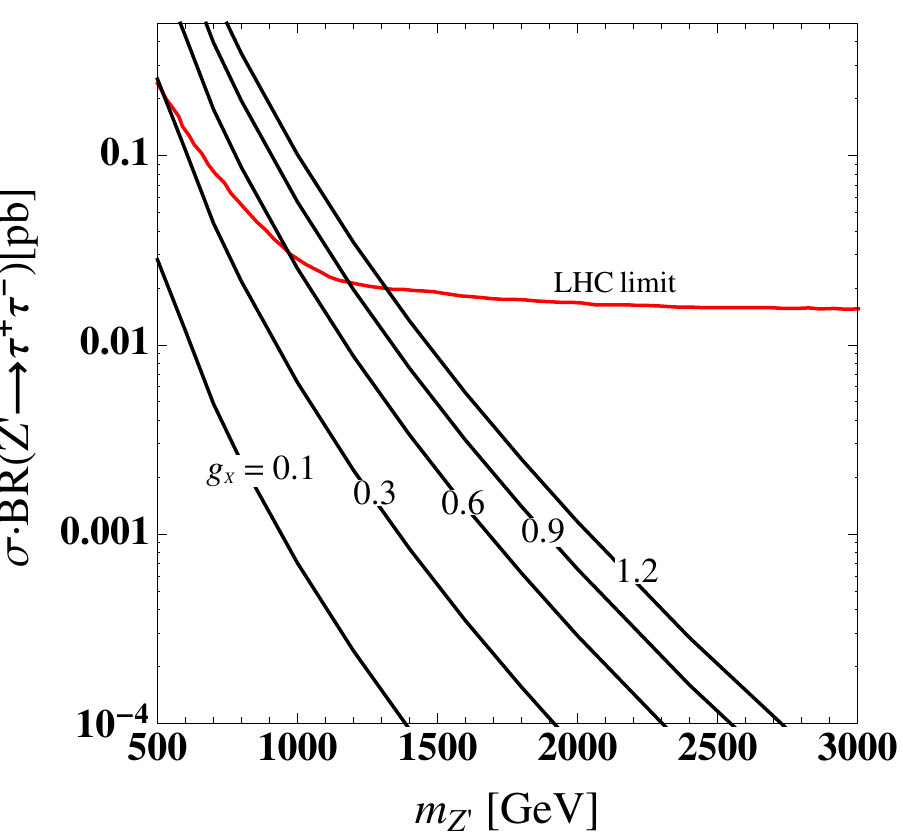} 
\caption{Left(right) plot: $\sigma(pp \to Z') BR(Z' \to \ell^+ \ell^- (\tau^+ \tau^-))$ with $\ell = e, \mu$ for several values of $g_X$ compared with LHC limit; from Refs.~\cite{Aaboud:2017buh} and \cite{Khachatryan:2016qkc} for $\ell^+ \ell^-$ and $\tau^+ \tau^-$ modes. }
\label{fig:LHC}
\end{center}
\end{figure}

Here we discuss  $Z'$ production at the LHC 13 TeV where $Z'$ can be produced via interaction 
in Eq.~(\ref{eq:int_Z'}), followed by decay modes of $Z' \to \mu^+ \mu^-$ and $Z' \to \tau^+ \tau^-$  
[Drell-Yan (DY) productions]. 
In this model $Z'$ mainly decays into $\tau^+ \tau^-$ model with $BR(Z' \to \tau^+ \tau^-) \sim 0.5$ and BR of $\mu^+ \mu^-$ mode is suppressed by factor of $1/16$.
The production cross section is estimated by CalcHEP~\cite{Belyaev:2012qa} using 
the CTEQ6 parton distribution functions (PDFs)
~\cite{Nadolsky:2008zw}.  
In Fig.~\ref{fig:LHC}, we show $\sigma (pp \to Z') BR(Z' \to \ell^+ \ell^-/\tau^+ \tau^-)$ ($\ell = e, \mu$) as a function of the $Z'$ mass for several values of $g_X$.
The cross sections are compared with constraints from LHC data; from Refs.~\cite{Aaboud:2017buh} and \cite{Khachatryan:2016qkc} for $\ell^+ \ell^-$ and $\tau^+ \tau^-$ modes.
We thus find that $\ell^+ \ell^-$ mode (mostly $\mu^+ \mu^-$) provides more strict bound although $BR(Z' \to \mu^+ \mu^-) : BR(Z' \to \tau^+ \tau^-) = 1: 16$. 
Here we set masses of $H$, $A$ and $H^\pm$ as 400 GeV and apply $\tan \beta = 10$ and 
$\cos(\alpha - \beta) = 0$  where the effects of the $Z'$ decays into scalar bosons are small. 
Also right-handed neutrinos and DM $\chi$ are taken to be heavier than $m_{Z'}/2$ so that $Z'$ 
does not decay into on-shell right-handed neutrinos and DM.

Our $Z'$ boson also decays into neutrinos with BR value of 
$BR(Z' \to \nu_\tau \bar \nu_\tau) = 16 BR(Z' \to \nu_\mu \bar \nu_\mu) \simeq 0.25$.
Thus we can also test our model by $pp \to Z' g \to \nu \bar \nu g$ process at the LHC experiments searching for signal with mono-jet plus missing transverse momentum.
The cross section of $pp \to Z' g \to \nu \bar \nu g$ process is, for example, $\sim 1$ fb with $g_X = 0.6$ and $m_{Z'} = 1500$ GeV estimated by CalcHEP with $p_T > 25$ GeV cut.
We thus need large integrated luminosity to analyze the signal~\cite{Aaboud:2017phn} and it will be tested in future LHC experiments. 


\subsection{Dark matter}
\begin{figure}[tb]
\begin{center}
\includegraphics[width=75mm]{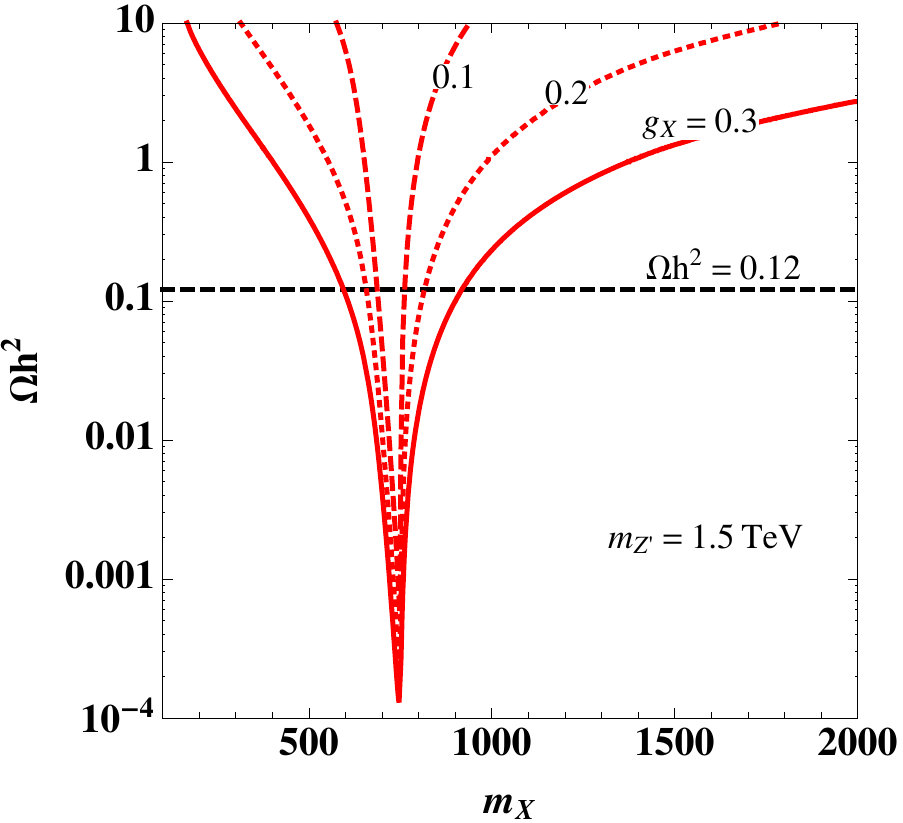} 
\caption{ Relic density of DM as function of DM masses  for different values of 
$U(1)_X$ gauge couplings, $g_X = 0.1, 0.2$ and $0.3$.  We have fixed $m_{Z'} = 1.5$ TeV. 
}
\label{fig:relic0}
\end{center}
\end{figure}

We consider a Dirac fermion $\chi$ as our DM candidate, and the relic density is determined 
by the DM annihilation process $\chi \bar \chi \to Z' \to f_{SM} \bar f_{SM}/HA/H^+H^-$ where 
$f_{SM}$ is a SM fermion and/or $\chi \bar \chi \to Z' Z'$ depending on kinematic condition.
Then we estimate relic density of our DM using {\tt micrOMEGAs 4.3.5}~\cite{Belanger:2014vza} 
implementing relevant interactions.
Fig.~\ref{fig:relic0} shows the relic density $\Omega h^2$ as a function of DM mass $m_X$ 
where we apply several values of $g_X$ and $m_{Z'} = 1.5$ TeV as reference values, and indicate 
observed $\Omega h^2$ value by horizontal dashed line~\cite{Aghanim:2018eyx}.
We see that the relic density drops at around $m_{Z'} \sim 2 m_X$ due to resonant enhancement 
of the annihilation cross section.

\begin{figure}[tb]
\begin{center}
\includegraphics[width=75mm]{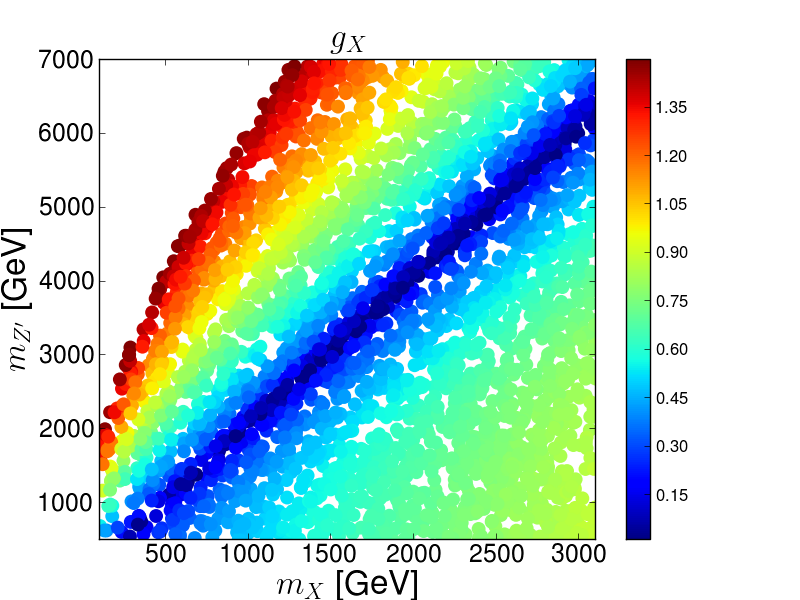} 
\includegraphics[width=75mm]{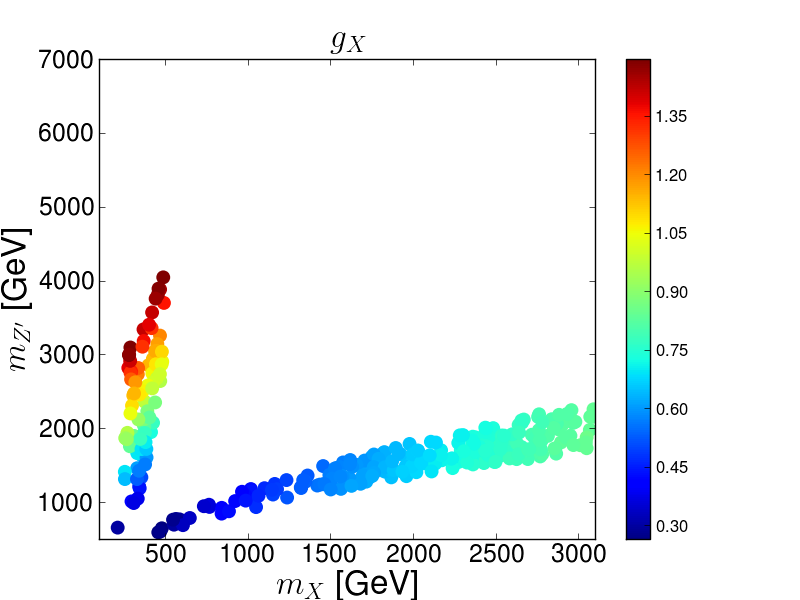}
\caption{
(Left): parameter region which accommodates the observed  DM relic density. 
(Right): parameter region which explains both DM relic density and $b \to  s \ell^+ \ell^-$ anomalies.}
\label{fig:relic}
\end{center}
\end{figure}

In addition, we scan parameters in the range of 
\begin{equation}
m_X \in [200, 3100] \ {\rm GeV}, \quad m_{Z'} \in [500, 7000] \ {\rm GeV}, \quad g_X \in [0.01, 1.5] ,
\end{equation}
with assuming that $\tan \beta = 10$ and $\cos (\alpha - \beta) = 0$ as reference values.  
We note that the effects of scalar bosons are subdominant.   
The left panel of Fig.~\ref{fig:relic} shows the parameter region 
which accommodates the observed relic density of DM, $\Omega h^2 = 0.1206 \pm 0.0063$, 
taking $3 \sigma$ range of observed value by  the Planck collaboration~\cite{Aghanim:2018eyx}.
Moreover the right panel of the figure indicates the region in which both observed relic density 
and $b \to  s \ell^+ \ell^-$ anomalies are explained within 2$\sigma$.
Notice that the allowed region with $m_{Z'} < m_X$ is partly excluded by or close to LHC constraint 
shown in Fig.~\ref{fig:LHC} and will be explored in future LHC experiments.
In addition DM-nucleon scattering cross section by $Z'$ exchange is suppressed by CKM factor 
and the allowed region is not constrained by the DM direct detection experiments.

Before closing this section we discuss possibility of indirect detection of our DM. 
In this model DM pair annihilates mainly through $\chi \bar \chi \to Z' \to \tau^+ \tau^-$ and/or $\chi \bar \chi \to Z' Z' \to 2 \tau^+ \tau^-$ and 
gamma-ray search gives the strongest constraint on the annihilation cross section by Fermi-LAT observation~\cite{Hoof:2018hyn, Fermi-LAT:2016uux}.
In our parameter region of $m_{Z'} > 500$ GeV, DM annihilation cross section explaining the relic density is well below the constraint for the $\tau^+ \tau^-$ dominant case~\cite{Fermi-LAT:2016uux, Hoof:2018hyn} unless there is large enhancement factor; constraint on cross section for four $\tau$ mode would be similar. 
Thus our model is safe from indirect detection cross section and will be tested with larger amount of data in future.

\section{ Summary and discussions}

We have discussed a flavor model based on $U(1)_{B_3 - x_\mu L_\mu - x_\tau L_\tau} 
(\equiv U(1)_X)$ gauge symmetry in which two Higgs doublet fields are introduced to obtain the 
observed CKM matrix.
Flavor changing $Z'$ interactions with the SM quarks are obtained after diagonalizing quark mass 
matrix, and $b \to  s \ell^+ \ell^-$ anomalies can be explained due to lepton flavor non-universal 
charge assignment when $x_\mu$ is taken to be negative value. 
Then we have considered minimal set up explaining $b \to  s \ell^+ \ell^-$ anomalies and 
generated neutrino mass matrix where two SM singlet scalar fields and Dirac fermionic DM 
candidate are introduced.

We have  computed the  $Z'$ contribution to the Wilson coefficient  $C_9^\mu$ relevant for 
$b\rightarrow s \mu^+ \mu^-$, as wel as neutrino mass matrices,  charged lepton flavor violations 
and  the $B_s$--$\bar B_s$ mixing, including the relevant experimental constraints. 
We have found that cancellation between $Z'$ and scalar bosons contributions 
to $B_s$--$\bar B_s$ is required to satisfy experimental constraint,   while explaining 
$b \to  s \ell^+ \ell^-$ anomalies. 
In addition, we have shown constraints from lepton flavor violation process 
$\mu \to e \gamma$ and future prospects for $\mu \to e$ conversion measurements.

Then collider physics regarding $Z'$ production at the LHC and relic density of DM are explored.
We have shown cross sections for the DY processes, $pp \to Z' \to \mu^+ \mu^- (\tau^+ \tau^-)$, 
where constraints on  the $\{ m_{Z'} , g_X \}$ parameter space dominantly come from the data of 
di-muon resonance search at the LHC.
The relic density of DM further constrains $\{ m_{Z'} , g_X \}$ parameter space since the relic 
density is determined by DM pair annihilation process via $Z'$ interactions.
The preferred parameter region can be further tested in future LHC experiments and observations 
for flavor physics such as LFVs.

\section*{Acknowledgments}
\vspace{0.5cm}
The work of CY was supported by the National Research Foundation of Korea(NRF) grant funded by
	the Korea government(MSIT), NRF-2017R1A2B4011946 and NRF-2017R1E1A1A01074699.

\section*{ Appendix: Yukawa interactions}

Here we summarize Yukawa interactions in two Higgs doublet sector which are taken from ref.~\cite{Crivellin:2015lwa}.
\begin{align}
\mathcal{L}_Y  = & - \bar u_L \left( \frac{\cos \alpha}{v \sin \beta} m_u^D - \frac{\cos (\alpha - \beta)}{\sqrt{2} \sin \beta} \tilde \xi^u \right) u_R h 
- \bar d_L \left( \frac{\cos \alpha}{v \sin \beta} m_d^D - \frac{\cos (\alpha - \beta)}{\sqrt{2} \sin \beta} \tilde \xi^d \right) d_R h  \nonumber \\
& - \bar u_L \left( \frac{\sin \alpha}{v \sin \beta} m_u^D - \frac{\sin (\alpha - \beta)}{\sqrt{2} \sin \beta} \tilde \xi^u \right) u_R H
- \bar d_L \left( \frac{\sin \alpha}{v \sin \beta} m_d^D - \frac{\sin (\alpha - \beta)}{\sqrt{2} \sin \beta} \tilde \xi^d \right) d_R H \nonumber \\
& - i \bar u_L \left( \frac{m_u^D}{v \tan \beta} - \frac{1}{\sqrt{2} \sin \beta} \tilde \xi^u \right) u_R A
+ i \bar d_L \left( \frac{m_d^D}{v \tan \beta} - \frac{1}{\sqrt{2} \sin \beta} \tilde \xi^d \right) d_R A \nonumber \\
& -  \left[ \bar u_R \left( \frac{\sqrt{2}}{v \tan \beta} m_u^D V - \frac{1}{\sin \beta} (\tilde \xi^u)^\dagger \right) d_L 
+ \bar u_L \left( \frac{\sqrt{2}}{v \tan \beta} V m_d^D - \frac{1}{\sin \beta} V \tilde \xi^d \right) d_R \right] H^+ \nonumber \\
& + h. c. \, ,
\label{Eq:YukawaH}
\end{align}
where flavor indices are omitted and the non-diagonal coupling matrices are defined as 
\begin{equation}
\tilde \xi^u = U_L^\dagger \frac{1}{\sqrt{2}} \begin{pmatrix} 0 & 0 & 0 \\ 0 & 0 & 0 \\ \tilde y^u_{31} & \tilde y^u_{32} & 0 \end{pmatrix} U_R, \quad 
\tilde \xi^d = D_L^\dagger \frac{1}{\sqrt{2}} \begin{pmatrix} 0 & 0 & \tilde y^d_{13} \\ 0 & 0 & \tilde y^d_{23} \\ 0 & 0 & 0 \end{pmatrix} D_R.
\end{equation}
Under the approximation $V \simeq D_L$ and $D_R \simeq {\bf 1}$, we obtain
\begin{equation}
\tilde \xi^d \simeq V^\dagger \xi^d \simeq \frac{\sqrt{2}}{\cos \beta} \frac{m_b}{v} 
\begin{pmatrix} 0 & 0 & - V_{td}^* V_{tb} \\ 0 & 0 & - V_{ts}^* V_{tb} \\ 0 & 0 & 1 - |V_{tb}|^2 \end{pmatrix}.
\end{equation}

\end{document}